\documentclass[twocolumn,showpacs,amsmath,amssymb,prb,superscriptaddress]{revtex4}
\usepackage{amsmath}
\usepackage{graphicx}% Include figure files
\usepackage{bm}% bold math
\begin{document}

\title{Nonequilibrium phenomena in multiple normal-superconducting tunnel
heterostructures}

\author{J. Voutilainen}
\affiliation{Low Temperature Laboratory, Helsinki University of
Technology, P.O. Box 2200, FIN-02015 HUT, Finland}
\author{T.\ T. Heikkil{\"a}}
\affiliation{Low Temperature Laboratory, Helsinki University of
Technology, P.O. Box 2200, FIN-02015 HUT, Finland}
\affiliation{Department of Physics and Astronomy, University of
Basel, Klingelbergstr. 82, CH-4056 Basel, Switzerland}
\author{N.\ B. Kopnin}
\affiliation{Low Temperature Laboratory, Helsinki University of
Technology, P.O. Box 2200, FIN-02015 HUT, Finland} \affiliation{L.
D. Landau Institute for Theoretical Physics, 117940 Moscow,
Russia}
\date{\today}

\begin{abstract}
Using the nonequilibrium theory of superconductivity with the
tunnel Hamiltonian, we consider a mesoscopic NISINISIN
heterostructure, i.e., a structure consisting of five intermittent
normal-metal (N) and superconducting (S) regions separated by
insulating tunnel barriers (I). Applying the bias voltage between
the outer normal electrodes one can drive the central N island
very far from equilibrium. Depending on the resistance ratio of
outer and inner tunnel junctions, one can realize either effective
electron cooling in the central N island or create highly
nonequilibrium energy distributions of electrons in both S and N
islands. These distributions exhibit multiple peaks at a distance
of integer multiples of the superconducting chemical potential. In
the latter case the superconducting gap in the S islands is
strongly suppressed as compared to its equilibrium value.
\end{abstract}

\maketitle

\section{Introduction}

Mesoscopic electronic applications typically rely on phenomena
which show best when the electrons in small wires are cooled to
very low temperatures, ideally to the range of 10 mK. In this
regime the crystal lattice is very weakly coupled to the electron
system, and electron cooling via the lattice becomes difficult. An
alternative approach is then to directly cool the electrons. This
can be achieved by placing superconducting (S) contacts via
insulating (I) barriers to the normal-metal (N) or superconducting
(S$^\prime$) wire whose electrons are to be cooled
\cite{nahum94,leivo96,GolubevVasenko}. By applying a voltage over
such SINIS/SIS$^\prime$IS coolers, it has been shown that one can
cool electrons well below 100 mK with these
structures\cite{Pekola2}, even when the lattice remains at a few
hundred mK, or to enhance the superconductivity in the middle
S$^\prime$ island
\cite{SCenhancement1,SCenhancement2,HeslingaKlapwijk}. Optimally,
such cooling should take the electron temperature to a few mK, a
limit which is hardly reached in mesoscopic systems via other
known means.

With this type of nonequilibrium cooling, the concept of the
electron temperature is not always well defined \cite{Pekola2},
but one has to rather describe the full electron energy
distribution function \cite{Pothier,Giazotto}. In this case
cooling corresponds to the sharpening of this distribution
function, essentially removing the high-energy excitations.

One of the features limiting the performance of such SINIS coolers
is the fact that the poor heat conductivity of the superconductors
makes them inefficient reservoirs \cite{Pekola1}. An additional
pair of normal-metal electrodes attached to superconducting
electrodes of a SINIS cooler can improve the relaxation and
enhance the cooling characteristics of the device
\cite{GolubevVasenko}. In this paper, we consider the effects of
extra N electrodes of this type attached to a generic SINIS
structure. The superconducting pieces are now assumed small enough
so that they can be driven out of equilibrium by applying a bias
voltage between the two normal electrodes. It is this arrangement
of a NISIN$^\prime$ISIN multiple heterostructure which is the
subject of the present study. Its novel feature as compared to the
traditional SINIS structure with bulk S electrodes is that
nonequilibrium is now induced in all inner islands of the
structure, which, in turn, strongly enhances nonequilibrium
effects both in the central N and in the adjacent S islands. The
resulting distributions in each island can be inspected by
transverse probe tunnel junctions\cite{Pothier,Pekola2}.

The microscopic nonequilibrium theory of double-barrier SINIS
junctions is based on time-dependent Keldysh Green function
formalism (see Ref.~\onlinecite{Brinkman} and references therein).
In the present paper we extend the theory such that nonequilibrium
distributions in superconducting islands are also allowed. The
major modification is that the chemical potentials of the two
superconducting islands cannot be chosen zero simultaneously since
these islands have different potentials determined by the bias
voltage. This is equivalent to a time dependence of the order
parameters imposed by different time-dependent order-parameter
phases. We restrict ourselves to a tunneling Hamiltonian
approximation which effectively makes the problem spatially
independent within each superconducting or normal island.

\section{Model}

\begin{figure}[t]
\centerline{\includegraphics[width=0.5\columnwidth]{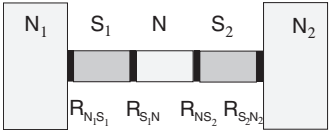}}
\caption{NISINISIN structure: The central normal island (N) is
placed between two superconducting islands (S) which are connected
to two outer N electrodes serving as external leads. Each
connection is realized through a tunnel contact with a resistance
$R_{ik}$.} \label{fig-5piece}
\end{figure}

The system we study is shown in Fig.~\ref{fig-5piece}. In
nonequilibrium, each of the islands has a separate energy
distribution and, according to our model based on the tunnel
Hamiltonian, they are independent of both coordinates and
directions of the momenta. This implies that each region is in a
diffusive regime when the momentum direction dependence is
averaged out within the first approximation in $\ell /L$ where
$\ell$ is the impurity mean free path and $L$ is the length of the
contacting region. In addition, one has to assume that the
intrinsic normal-state resistance $r$ of each island is much
smaller than any of the tunnel resistances $R$ to satisfy the
condition that the potential variation inside each region is
smaller than its drop across the tunnel barrier.

%%%%%%%%%%%%%%%%%%%%%%%%%%%%%%%%%%%%%%%%%%%%%%%%%%%%%%%%%%%%%%%%%%%%%
%%%%%%%%%%%%%%%%%%%%%%%%%%%%%%%%%%%%%%%%%%%%%%%%%%%%%%%%%%%%%%%%%%%%%%

A characteristic rate for tunneling from region 2 into region 1 to
be defined later is $ \eta _{12}= \left[4\nu _{1}e^2
\Omega_1R\right]^{-1}$, where $\nu_1$ is the normal-state density
of states (DOS) at the Fermi level in region 1 and $\Omega_1$ is
the volume of conductor 1. Competing with the inelastic
relaxation, this rate determines whether the injection of new
quasiparticles into the system is fast enough to drive the system
out of equilibrium. The inelastic processes which uphold the
thermal Fermi distribution can be neglected if $\eta \gg 1/\tau
_{\rm inel}$ which is equivalent to
\begin{equation}
\nu e^2\Omega R/\tau_{\rm inel} \sim \left(L\ell /\ell ^2 _{\rm
inel}\right)\Gamma \ll 1 \label{small-inel1}
\end{equation}
where $ \Gamma =(N_mR/R_Q)(\bar{A} /A)$. Here $\bar{A}$ stands for
the average cross-section area of the conductor, $L$ is its length
such that $\Omega =\bar{A}L$, $A$ is the area of the junction,
$\ell_{\rm inel}=\sqrt{D\tau_{\rm inel}}$, and $D=v_F\ell /3$ is
the diffusion coefficient. We also introduce the quantum
resistance $R_Q=\pi \hbar /2e^2$, and the number of conducting
modes $N_m\sim p_F^2A/\hbar ^2$ in the area of the junction.

In our work, we assume that the electron energy distribution
within each island is homogeneous, which also implies a
homogeneous potential and a constant effective temperature.
Analysis of kinetic equations shows that this is achieved if $
r\ll R $ or
\[
\sigma \bar{A}/L\sim \nu De^2 \bar{A}/L \gg 1/R
\]
which gives
\[
DL^{-2}\gg (\nu e^2R\Omega )^{-1}\sim \eta \ .
\]
This puts a restriction $L\ll \ell \Gamma$. The limit $\ell \ll L$
holds if $\Gamma \gg 1$. Our model is thus applicable provided
$1\ll L/\ell \ll \Gamma$. Condition (\ref{small-inel1}) of small
inelastic interaction reads $ \ell_{\rm inel}^2\gg \Gamma L\ell $.

\section{Formalism}

\subsection{Transport equation}

We use the standard Keldysh Green function formalism, see for
example Ref.~\onlinecite{KopninBook}. We denote the Nambu-space
matrix operator
\[
\hat G^{-1} =\hat \tau _3 \frac{\partial }{\partial \tau}
-\frac{\nabla ^2}{2m} -\mu  +\hat H \, , \; \hat H
=\left(\begin{array}{cc} 0& -\Delta \\ \Delta ^* &0
\end{array}\right).
\]
It acts on matrix Green functions
\[
\hat G =\left(\begin{array}{cc} G&F\\-F^\dagger &\bar
G\end{array}\right) \ ,
\]
where $\hat G$ stands for either retarded (advanced), $\hat
G^{R(A)}$, or Keldysh, $\hat G^K$, Green function. For tunnelling
between two superconductors (or normal metals) 1 and 2, the self
energy in region 1 is $ \hat \Sigma _T(1)=i\eta _{12}\hat g(2) $
provided the semi-classical Green functions
\[
\hat g=\int \frac{d\xi _p}{\pi i} \hat G\left({\bf p}+{\bf k}/2,
{\bf p}-{\bf k}/2\right),
\]
integrated over the energy $\xi _p=p^2/2m -E_F$, are independent
of the directions of momentum ${\bf p}$. The tunneling rate is
defined as
\begin{equation}
\eta _{12}= \pi \nu _{2}\Omega _1^{-1}\left< |T_{{\bf p},{\bf
q}}|^2\right>=\left[4\nu _{1}e^2 \Omega _1R_{12}\right]^{-1} \ .
\label{eta}
\end{equation}
Here $T_{{\bf p},{\bf q}}$ is the tunnel matrix element and
$R_{12}$ is the tunnel resistance of the contact between regions 1
and 2. Note that the self-energy in region 2 is $\hat \Sigma
_T(2)=i\eta _{21}\hat g(1)$ where $ \eta _{21}= \pi \nu _{1}\Omega
_2^{-1}\left< |T_{{\bf p},{\bf q}}|^2\right>=\left[4\nu _{2}e^2
\Omega _2R_{12}\right]^{-1}$. If region 1 has tunnel contacts to
two other regions 2 and 3, the self energy is a sum
\[
\hat \Sigma _T(1)=i\eta _{12}\hat g(2)+i\eta _{13}\hat g(3)
\]
where $ \eta_{12}=\left[4\nu_{1}e^2\Omega_1R_{12}\right]^{-1}$ and
$ \eta_{13}=\left[4\nu_{1}e^2\Omega_1R_{13}\right]^{-1}$.

Including the tunnelling self-energy into the total self-energy
that contains both elastic and inelastic processes we can write
equations for the retarded, advanced, and Keldysh Green functions:
\begin{eqnarray*}
\left(\hat G^{-1} -\hat \Sigma ^{R(A)}\right)\circ \hat G^{R(A)}
&=&\hat 1 \delta(x_1-x_2) \ , \\
\left(\hat G^{-1} -\hat \Sigma ^{R}\right)\circ \hat G^{K} -\hat
\Sigma ^{K}\circ \hat G^{A}&=&0 \ .
\end{eqnarray*}
The product $\hat \Sigma ^{K}\circ \hat G^{A}$ is a convolution
over frequencies and momenta of the type
\[
\hat \Sigma ^{K}\circ \hat G^{A}=\int \hat \Sigma
^{K}_{\epsilon_1,\epsilon} \hat G^{A}_{\epsilon, \epsilon_2}\,
\frac{d\epsilon}{2\pi} \ .
\]
Applying the inverse operator to $\hat G^{R(A)}$ and $\hat G^{K}$
from the right and subtracting the obtained equations from the
equations above we find the transport-like equations
\begin{eqnarray}
{\bf v}_F{\bf k}\hat G^K_{\epsilon _1,\epsilon _2}-\epsilon _1\hat
\tau _3\hat G^K_{\epsilon _1,\epsilon _2}+\hat G^K_{\epsilon
_1,\epsilon _2}\epsilon _2\hat \tau _3 &&\nonumber \\
+\hat H \circ \hat G^K-\hat G^K\circ \hat H &=&\hat {\cal
I}^K_{\epsilon _1,\epsilon _2}\ , \label{eq-Keldysh}\\
{\bf v}_F{\bf k}\hat G^{R(A)}_{\epsilon _1,\epsilon _2}-\epsilon
_1\hat \tau _3\hat G^{R(A)}_{\epsilon _1,\epsilon _2}+\hat
G^{R(A)}_{\epsilon _1,\epsilon _2}\epsilon _2\hat \tau _3 &&
\nonumber \\
+\hat H \circ \hat G^{R(A)}-\hat G^{R(A)}\circ \hat H &=&\hat
{\cal I} ^{R(A)}_{\epsilon _1,\epsilon _2} \ ,
\label{eq-Eilenberger}
\end{eqnarray}
where the collision integrals in each region are
\begin{eqnarray*}
\hat {\cal I}^K_{\epsilon _1,\epsilon _2}&=&\hat \Sigma ^R \circ
\hat G^K-\hat G^K\circ \hat \Sigma ^A - \hat G^R\circ \hat \Sigma
^K+\hat \Sigma ^K\circ \hat G^A \ , \\
\hat {\cal I}^{R(A)}_{\epsilon _1,\epsilon _2}&=& \hat \Sigma
^{R(A)}\circ \hat G^{R(A)}-\hat G^{R(A)}\circ \hat \Sigma ^{R(A)}
\ .
\end{eqnarray*}
We also define
\[
\hat I^K_{\epsilon _1,\epsilon _2}\equiv \left(\begin{array}{cc}
I_1 & I_2 \\ -I_2^\dagger & \bar I_1
\end{array} \right) =\int \frac{d\xi _{p}}{\pi i}\hat {\cal
I}^K_{\epsilon _1,\epsilon _2} \ .
\]
Finally, the tunnel collision integral for region 1 in contact
with region 2 takes the form
\begin{eqnarray}
\hat I^K_{T}(1)&=&i\eta _{12} \left[ \hat g^R (2) \circ \hat
g^K(1)-\hat g^R (1) \circ \hat g^K (2)\right. \nonumber \\
&&\left. +\hat g^K (2) \circ \hat g^A (1)-\hat g^K(1) \circ \hat
g^A (2)\right]\ . \label{IKT-gen}
\end{eqnarray}

If the distribution function is time independent, the Keldysh
Green function has the form
\begin{equation}
\hat g^K_{\epsilon _1 , \epsilon _2}=\hat g^R_{\epsilon _1 ,
\epsilon _2}\left(f_{1, \epsilon _2}+\hat \tau _3 f_{2,\epsilon
_2}\right)-\left(f_{1, \epsilon _1}+\hat \tau _3 f_{2,\epsilon
_1}\right)\hat g^A_{\epsilon _1 , \epsilon _2}, \label{Keldysh}
\end{equation}
where the odd- and even-in-$\epsilon$ distribution functions $f_1$
and $f_2$ are defined such that
\[
f_{1,\epsilon }=-n(\epsilon )+n(-\epsilon)\, , \;
f_{2,\epsilon}=1-n(-\epsilon)-n(\epsilon)\ .
\]
The actual quasiparticle energy distribution function $n$
satisfies $f_{1,\epsilon }+f_{2,\epsilon }=1-2n(\epsilon)$.

For spatially uniform distributions the kinetic equations in each
conducting region is obtained after averaging
Eqs.~(\ref{eq-Keldysh}), (\ref{eq-Eilenberger}) over the momentum
directions. The elastic collision integrals average out, and the
kinetic equations for diagonal in $\epsilon$ components become
\begin{eqnarray}
{\rm Tr}\left(\hat M_{\epsilon ,\epsilon}\right)& =&{\rm Tr}\left[
\hat I^K_{\epsilon, \epsilon} -\left( \hat I^R_{\epsilon,
\epsilon}- \hat I^A_{\epsilon,
\epsilon}\right)f_{1,\epsilon}\right] \ ,
\label{kineq-1}\\
{\rm Tr}\left(\hat \tau_3\hat M_{\epsilon ,\epsilon}\right)&=&{\rm
Tr}\left(\hat \tau_3\left[\hat I^K -\left( \hat I^R_{\epsilon,
\epsilon}- \hat I^A_{\epsilon,
\epsilon}\right)f_{1,\epsilon}\right]\right) \ . \label{kineq-2}
\end{eqnarray}
The collision integrals here include only tunnel and inelastic
contributions $ \hat I =\hat I_T + \hat I_{\rm inel} $, where
inelastic processes include electron-phonon and electron-electron
interactions, $\hat I_{\rm inel}=\hat I_{\rm e-ph}+\hat I_{\rm
e-e}$. The matrix $\hat M$ is
\begin{eqnarray*}
\hat M &=&\left(\hat H\circ \hat g^K-\hat g^K\circ\hat H\right) \\
&&-\left( \hat H\circ \hat g^R-\hat g^R\circ \hat H\right)f_1+ f_1
\left(\hat H\circ \hat g^A-\hat g^A\circ \hat H\right) \ .
\end{eqnarray*}

\subsection{Charge and energy tunnel currents}

After taking the trace and integrating over energy and momentum,
Eq.~(\ref{eq-Keldysh}) in region 1 yields
\begin{eqnarray*}
-\frac{\partial}{\partial t}\left(\nu_{1}\int
\frac{d\epsilon}{2}\int \frac{d\xi _1}{\pi i}\frac{d\Omega_{\bf
p}}{4\pi} {\rm Tr}\, \hat G^K \right) \\
= -i\nu _{1}\int \frac{d\epsilon}{2}\frac{d\Omega_{\bf p}}{4\pi}
{\rm Tr}\,\left[\hat \tau _3\hat I^K(1)\right]\ ,
\end{eqnarray*}
which is the conservation of charge $Q(1)=eN(1)\Omega_1$ in the
volume $\Omega_1$ of region 1:
\begin{equation}
\frac{\partial Q(1)}{\partial t}=I(1) \ . \label{part-cons}
\end{equation}
Here $N(1)$ is the particle density in region 1. The current that
flows into region 1 is
\begin{equation}
I(1)=-ie\nu _{F1}\Omega_1 \int
\frac{d\epsilon}{4}\frac{d\Omega_{\bf p}}{4\pi} {\rm
Tr}\,\left[\hat \tau _3\hat I^K_T(1) \right] \ .
\label{charge-cons}
\end{equation}
The inelastic collision integral drops out because it conserves
the particle number.

Multiplying Eq. (\ref{eq-Keldysh}) by $\epsilon$ and taking the
trace of it one can obtain the balance of the energy ${\cal
E}=\Omega E$ in the form
\[
\frac{\partial {\cal E}(1)}{\partial t} = I_{\cal E}(1).
\]
Here the energy density is
\[
E =-\int \epsilon {\rm Tr}\, \left(\hat \tau _3\hat G^K_{\epsilon
_+,\epsilon _-}\right)\frac{d\epsilon }{4\pi i}\frac{d^3p}{(2\pi
)^3} +\frac{|\Delta |^2}{|g|}-Ne\varphi
\]
and the energy current into region 1 is
\begin{equation}
I_{\cal E}(1)=-i\nu _{1}\Omega_1 \int
\frac{d\epsilon}{4}\frac{d\Omega_{\bf p}}{4\pi} {\rm
Tr}\,\left[\left(\epsilon -e\varphi _1 \hat \tau _3 \right)\hat
I^K(1) \right] \ . \label{energy-curr}
\end{equation}

The collision integral in Eq. (\ref{energy-curr}) contains both
tunnel and electron-phonon contributions, $I^K(1)=I^K_{\rm e-ph}
(1)+I^K_T(1)$. The electron-electron interactions drop out because
of the energy conservation. The energy flow into region 1 can be
separated into two parts. One part containing $I^K_{\rm e-ph}$ is
the energy exchange with the heat bath (phonons). The other part
contains the tunnel contribution $I^K_T(1)$ and is the energy
current into region 1 through the tunnel contact.

%%%%%%%%%%%%%%%%%

\section{Kinetic equations
in hybrid structures}

\subsection{Nonzero superconducting chemical potential}

If a hybrid structure containing more than one superconducting
island is voltage biased, at least one of the superconductors will
have a nonzero chemical potential with a time dependent phase of
the order parameter. If the external voltage is constant and the
resulting state is stationary, the phase is linear in time, $\chi
=2\mu _S t$, while the magnitude of the order parameter is time
independent. Consider two sets of regular Green functions. One set
of functions $\hat g^{R(A)}$ is taken in the basis where the order
parameter phase varies in time. The other set belongs to the same
order parameter magnitude but has a phase constant in time (zero
for simplicity). However, the energies of the Green-functions
$g^{R(A)}$, $\bar g^{R(A)}$, $f^{R(A)}$, and $f^{\dagger R(A)}$
are shifted by different amounts proportional to the chemical
potential in the previous representation. The relations between
these two sets, the Green functions $g^{R(A)}_{\epsilon, \epsilon
^\prime}$ and $f^{R(A)}_{\epsilon, \epsilon ^\prime}$ and the
steady-state functions with shifted energies, can be established
from the Eilenberger equations. Since $\Delta =|\Delta|e^{+i2\mu
_St}$, i.e.,
\[
\Delta _\omega =|\Delta |2\pi \delta(\omega+2\mu _S) \, , \;
\Delta ^*_\omega =|\Delta |2\pi \delta(\omega -2\mu _S) \ ,
\]
one observes that Eqs.~(\ref{eq-Eilenberger}) are satisfied by
\begin{equation}
g^{R(A)}_{\epsilon ,\epsilon ^\prime}=g^{R(A)}_{\epsilon
+\mu_S}2\pi \delta (\epsilon -\epsilon ^\prime ) \, , \; \bar
g^{R(A)}_{\epsilon ,\epsilon ^\prime}=\bar g^{R(A)}_{\epsilon
-\mu_S}2\pi \delta (\epsilon -\epsilon ^\prime ) \label{g-mu}
\end{equation}
and
\begin{eqnarray}
f^{\dagger R(A)} _{\epsilon,\epsilon^\prime}&=&f^{\dagger R(A)}
_{\epsilon -\mu _S} 2\pi \delta(\epsilon -\epsilon ^\prime -2\mu
_S) \ , \nonumber \\
f^{R(A)} _{\epsilon,\epsilon^\prime} &=& f^{R(A)} _{\epsilon +\mu
_S} 2\pi \delta(\epsilon -\epsilon ^\prime +2\mu _S) \ .
\label{f-mu}
\end{eqnarray}
The functions $g^{R(A)}_\epsilon =-\bar g^{R(A)}_\epsilon$ and
$f^{R(A)}_\epsilon =f^{\dagger R(A)}_\epsilon$ satisfy the
steady-state Eilenberger equation
\begin{equation}
-2\epsilon f^{R(A)}_\epsilon +2|\Delta |g^{R(A)}_\epsilon
=I_{2,\epsilon}^{R(A)} \label{Eilen-eq1}
\end{equation}
supplemented with the normalization
\begin{equation}
\left[g^{R(A)}_\epsilon\right]^2-\left[f^{R(A)}_\epsilon
\right]^2=1 \ . \label{Eilen-eq2}
\end{equation}

Therefore, we find from Eq. (\ref{Keldysh}) for a nonzero chemical
potential
\begin{eqnarray}
g^K_{\epsilon, \epsilon ^\prime}&=& \left[g^{R}_{\epsilon +\mu
_s}- g^{A}_{\epsilon +\mu _s}\right]
\left(f_{1,\epsilon}+f_{2,\epsilon}\right)2\pi
\delta (\epsilon -\epsilon ^\prime)  \ , \quad \label{gK} \\
-\bar g^K_{\epsilon, \epsilon ^\prime}&=& \left[ g^{R}_{\epsilon
-\mu _s}-  g^{A}_{\epsilon -\mu _s}\right]
\left(f_{1,\epsilon}-f_{2,\epsilon}\right)2\pi \delta (\epsilon
-\epsilon ^\prime) \ . \quad \label{bargK}
\end{eqnarray}
The off-diagonal Keldysh Green function in Eq. (\ref{Keldysh})
becomes
\begin{eqnarray}
f^K_{\epsilon , \epsilon -\omega} &=&\left[f^R_{\epsilon
+\mu_S}(f_{1,\epsilon
+2\mu_S}-f_{2,\epsilon +2\mu_S})\right. \nonumber \\
&&\left. -(f_{1,\epsilon}+f_{2, \epsilon})f^A_{\epsilon
+\mu_S}\right]\, 2\pi \delta(\omega+2\mu_s) \ . \label{fK}
\end{eqnarray}

%%%%%%%%%%%%%%%%%%%%%%
\subsection{Tunnel collision integrals}

Assume that region 1 is a superconductor and region 2 is a normal
metal. With Eqs.~(\ref{g-mu}), (\ref{gK}), and (\ref{bargK}), the
matrix elements of the tunnel collision integral
Eq.~(\ref{IKT-gen}) in the superconducting region $S$ give
\begin{widetext}
\begin{eqnarray}
{\rm Tr}\left[\hat I^K_T(S)\right]&=&4i\eta_{SN} \left\{ \left[
g_{\epsilon +\mu_S}(S)+g_{\epsilon -\mu_S}(S)\right]
\left[f_{1,\epsilon}(S)-f_{1,\epsilon }(N)\right] +\left[
g_{\epsilon +\mu _S} (S)-g_{\epsilon -\mu_S}(S)\right]
\left[f_{2,\epsilon}(S)-f_{2,\epsilon}(N)\right]\right\}  ,\qquad
\label{I-trace1} \\
{\rm Tr}\left[\hat \tau _3\hat I^K_T(S)\right]&=&4i\eta_{SN}
\left\{ \left[ g_{\epsilon +\mu_S}(S)-g_{\epsilon
-\mu_S}(S)\right] \left[f_{1,\epsilon}(S)-f_{1,\epsilon
}(N)\right]+\left[ g_{\epsilon +\mu _S} (S)+ g_{\epsilon
-\mu_S}(S)\right]
\left[f_{2,\epsilon}(S)-f_{2,\epsilon}(N)\right]\right\}  . \qquad
\label{I-trace2}
\end{eqnarray}
Components of the retarded and advanced tunnel collision integrals
$\hat I^{R(A)}$ are
\[
I^{R(A)}_{2}(S)=\pm i\eta_{SN} f^{R(A)}(S)\, , \; -I^{\dagger
R(A)}_{2}(S)=\pm i\eta_{SN} f^{\dagger R(A)}(S)\, ,\;
I^{R(A)}_{1}(S)=\bar I^{R(A)}_{1}(S)=0 \ .
\]
Both Keldysh and R(A) tunnel collision integrals in the normal
region N are coupled to those in the region S by $
\eta_{SN}I(N)=-\eta_{NS}I(S) $.

Using Eqs.~(\ref{f-mu}) and (\ref{fK}), we find
\begin{eqnarray}
{\rm Tr}\left(\hat M_{\epsilon, \epsilon}\right)&=&2|\Delta
|F_{\epsilon +\mu_S} \left(f_{1,\epsilon}-f_{1,\epsilon
+2\mu_S}+f_{2, \epsilon}+ f_{2,\epsilon +2\mu_S}\right) -2|\Delta
|F^R_{\epsilon -\mu_S} \left(f_{1,\epsilon
-2\mu_S}-f_{1,\epsilon}+f_{2,
\epsilon}+ f_{2,\epsilon -2\mu_S}\right) , \quad \label{M-trace1} \\
{\rm Tr}\left(\hat \tau _3\hat M_{\epsilon,
\epsilon}\right)&=&2|\Delta |F_{\epsilon +\mu_S}
\left(f_{1,\epsilon}-f_{1,\epsilon +2\mu_S}+f_{2, \epsilon}+
f_{2,\epsilon +2\mu_S}\right) +2|\Delta |F_{\epsilon -\mu_S}
\left(f_{1,\epsilon -2\mu_S}-f_{1,\epsilon}+f_{2, \epsilon}+
f_{2,\epsilon -2\mu_S}\right) . \quad \label{M-trace2}
\end{eqnarray}
\end{widetext}
After inserting Eqs.~(\ref{I-trace1}), (\ref{I-trace2}) and
(\ref{M-trace1}), (\ref{M-trace2}) into Eqs.~(\ref{kineq-1}) and
(\ref{kineq-2}) they yield the final kinetic equations to be
solved for the distribution functions.

In the equations above, we introduce $ g_\epsilon
=\left(g^R_\epsilon -g^A_\epsilon \right)/2\equiv {\rm Re}\,
g^R_\epsilon $ and also $ F_\epsilon =\left( f^R_\epsilon
+f^A_\epsilon \right)/2\equiv i\, {\rm Im}\, f^R_\epsilon  $. The
functions $g_\epsilon$ and $F_\epsilon $ are even in $\epsilon$.
With the account of tunnel and electron-phonon interactions, they
can be found from the steady-state Eilenberger equations
(\ref{Eilen-eq1}), (\ref{Eilen-eq2}). These equations can be more
easily solved when the inelastic interaction is small. If region 2
is normal, one finds
\begin{eqnarray}
g^{R(A)}_\epsilon &=&\pm \frac{\epsilon \pm
i\gamma}{\sqrt{\left(\epsilon
\pm i\gamma \right)^2 -|\Delta |^2}} \, , \label{gRA} \\
f^{R(A)}_\epsilon &=&\pm \frac{|\Delta |}{\sqrt{\left(\epsilon \pm
i\gamma \right)^2 -|\Delta |^2}} \ , \label{fRA}
\end{eqnarray}
where $\gamma =\eta_{SN}$.

For a N$_1$IS$_1$INIS$_2$IN$_2$ structure, each conductor S$_1$,
N, or S$_2$ has tunnel contacts with two other conductors, thus
the tunnel integrals are sums of the contributions from two
regions. For example, if the inelastic interaction is small, the
depairing rate in Eqs.~(\ref{gRA}) and (\ref{fRA}) is $
\gamma(S_i) =\eta _{S_iN}+\eta_{S_iN_i} $.

%%%%%%%%%%%%%%
\subsection{Current conservation}

For a NS junction, the tunnel current Eq.~(\ref{charge-cons}) into
the superconductor S from a normal region N or N$_{1,2}$ is
\begin{widetext}
\begin{eqnarray}
I(N\rightarrow S)&=&\frac{1}{4eR_{SN}}\int_{-\infty}^{\infty}
d\epsilon \, g_{\epsilon}(S) \left[f_{1,\epsilon
-\mu_S}(S)-f_{1,\epsilon +\mu_S}(S)-f_{1,\epsilon -\mu _S}(N)
+f_{1,\epsilon +\mu _S}(N)\right. \nonumber \\
&&+\left. f_{2,\epsilon -\mu_S }(S)+ f_{2,\epsilon +\mu_S
}(S)-f_{2,\epsilon -\mu_S }(N) -f_{2,\epsilon +\mu_S }(N)\right] \
. \label{e-current/NIS}
\end{eqnarray}
Electric currents through both junctions are balanced when
$I(N\rightarrow S_i)+I(N_i\rightarrow S_i)=0$.

The energy current flowing into the superconducting island through
each junction has the form
\begin{eqnarray}
I_{\cal E}(N\rightarrow
S)&=&\frac{1}{4e^2R_{NS}}\int_{-\infty}^{\infty} d\epsilon\,
\epsilon \, g_{\epsilon}(S) \left(\left[f_{1,\epsilon
-\mu_{S}}(S)+f_{1,\epsilon +\mu_{S}}(S)-f_{1,\epsilon -\mu
_{S}}(N)
-f_{1,\epsilon +\mu _{S}}(N)\right] \right. \nonumber \\
&&+\left. \left[f_{2,\epsilon -\mu_{S} }(S)- f_{2,\epsilon
+\mu_{S} }(S)-f_{2,\epsilon -\mu_{S} }(N) +f_{2,\epsilon +\mu_{S}
}(N)\right]\right)-I(N\rightarrow S)[\mu _{S}/e+\varphi _{S}] \ .
\label{IENS}
\end{eqnarray}
\end{widetext}
The energy conservation follows from the kinetic equations and is
therefore an abundant condition. However, in the quasi-equilibrium
limit, when the electron-electron interaction in each island
dominates over the tunnel injection, the distribution functions
are nearly thermal with certain temperatures specific for each
region. In this case, the kinetic equations do not need to be
solved explicitly, but the temperatures are found by requiring the
conservation of the energy current.

The energy current $I_{\cal E}(S\rightarrow N)$ flowing into the
central normal island through each junction is obtained from
Eq.~(\ref{IENS}) by changing the sign and replacing $\varphi _{S}$
with $\varphi _{N}$ so that the difference between the energy
currents from N to S calculated in two opposite directions is $
I_{\cal E}(N\rightarrow S)+I_{\cal E}(S\rightarrow
N)=I(N\rightarrow S)V_{NS}, $ where $V_{NS}=\varphi_N-\varphi _S $
is the voltage across the junction. This difference of energy
currents is compensated by the work produced by the voltage source
and does not lead to the change in the energy of the
superconductor. When the energy currents into a superconducting
island are balanced, the terms with $\mu _{S}/e+\varphi _{S}$
cancel due to the electric current conservation. In the normal
island, however, only the terms with $\varphi _N$ drop out since
the chemical potentials $\mu_{S}$ on both sides of the island are
different.

%%%%%%%%%%%%%%%%%%%

\subsection{Self-consistency equation and charge neutrality}

The magnitude of the order parameter is found from Eq.~(\ref{fK}).
It is a real quantity, therefore,
\begin{eqnarray}
\frac{|\Delta|}{\lambda} = \int_{-E_c}^{E_c} \left({\rm Re}\,
f^R_{\epsilon }\right)\left[f_{1,\epsilon +\mu_S}+f_{1,\epsilon
-\mu_S} \right. \nonumber \\
\left. -f_{2,\epsilon +\mu_S}+f_{2, \epsilon-\mu_S}\right]\,
\frac{d\epsilon}{4} \label{eq-Delta}
\end{eqnarray}
where $E_c$ is the BCS cut-off energy. The self-consistency of the
order parameter requires vanishing of its imaginary part:
\begin{equation}
\int_{-\infty}^{\infty} F_{\epsilon }\left(f_{1,\epsilon +\mu_S}
-f_{1,\epsilon -\mu_S}-f_{2,\epsilon +\mu_S}-f_{2, \epsilon
-\mu_S}\right)\, d\epsilon =0 \ , \label{condition1}
\end{equation}
which results in
\begin{equation}
\int {\rm Tr}\hat M \, d\epsilon =\int {\rm Tr}\left(\hat
\tau_3\hat M\right)\, d\epsilon =0 \ . \label{zerotrace}
\end{equation}

Equation (\ref{condition1}) or the second condition in
Eq.~(\ref{zerotrace}) together with the kinetic equation
Eq.~(\ref{kineq-2}) is equivalent to $ \int {\rm Tr}\left(\hat
\tau _3\hat I^K\right)\, d\epsilon =0 $. Since the integrals over
the inelastic electron-phonon and electron-electron collision
integrals vanish, this is the condition of current conservation:
\begin{equation}
\int {\rm Tr}\left(\hat \tau _3\hat I^K_T\right)\, d\epsilon =0\ ,
\label{condition2}
\end{equation}
which implies through Eq.~(\ref{charge-cons}) that the sum of
currents flowing into an island is zero.

Charge density can be obtained by calculating $ {\rm Tr}\hat g^K$
from Eqs.~(\ref{gK}), (\ref{bargK}). Making shift of $\epsilon$ in
the integral we obtain
\begin{eqnarray*}
&N&=N_0-2\nu \left(e\varphi +\mu _S\right)\nonumber \\
&+&\nu \int_{-\infty}^{\infty} g_\epsilon \left( f_{1,\epsilon
+\mu_S} -f_{1,\epsilon -\mu_S} -f_{2,\epsilon +\mu_S}-f_{2,
\epsilon -\mu_S}\right)\frac{d\epsilon }{2}\ .
\end{eqnarray*}
Here we take into account the divergence of the integral for large
$\epsilon$. Charge neutrality requires that $N=N_0$. The electric
potential $\varphi_S$ of a superconductor is thus coupled to the
chemical potential through
\begin{equation}
e\Phi =\int_{-\infty}^{\infty} g_\epsilon \left(
f_{1,\epsilon+\mu_S} -f_{1,\epsilon -\mu_S}- f_{2,\epsilon
+\mu_S}-f_{2,\epsilon -\mu_S}\right)\frac{d\epsilon}{4} \ ,
\label{charge1}
\end{equation}
where $ e\Phi =\mu_S+e\varphi_S$.

To summarize, for a system of $n$ superconducting and $m=n+1$
normal pieces we have $n+m$ pairs of the distribution functions
$f_1$ and $f_2$ and $n$ parameters $\mu_S$. For these unknowns, we
have $2(n+m)$ functional kinetic equations Eqs.~(\ref{kineq-1}),
(\ref{kineq-2}) and $n$ conditions Eq.~(\ref{condition2}) of
current conservation for each superconducting island. Note that
the current balance of the type of Eq.~(\ref{condition2}) in each
normal island is satisfied automatically due to the kinetic
equations that have $\hat M\equiv 0$ in any normal conductor.

\section{Results}

To further simplify our model, we limit the discussion to a case
symmetric with respect to the central normal island, i.e.
$R_{N_1S_1}=R_{N_2S_2}\equiv R_{\rm out}$,
$R_{NS_1}=R_{NS_2}\equiv R_{\rm in}$, $|\Delta _1|=|\Delta
_2|\equiv \Delta $, and $\varphi_{N_1} =-\varphi_{N_2}$,
$\mu_{S_1}=-\mu_{S_2}\equiv \mu _S$. Therefore $f_1(N_1)=f_1(N_2)$
and $f_2(N_1)=-f_2(N_2)$. Moreover, when applied to kinetic
equations in the normal region N, the symmetry yields $
f_2(S_1)=-f_2(S_2)\equiv f_2(S)\, , \; f_1(S_1)=f_1(S_2)\equiv
f_1(S)$ and $f_2(N) =0$.

\subsection{Nonequilibrium state}

\subsubsection{Distribution functions}

%%%%%%%%%%%%%%%%%%%%%%%%
\begin{figure}[t]
\centerline{\includegraphics[width=1.00\linewidth]{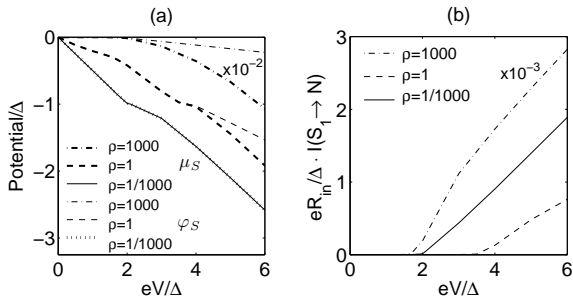}}
\caption{(a) Chemical and electrostatic potentials of the
superconducting island as a function of bias voltage. (b) $I-V$
curves for different $\rho$. For clarity, in the case $\rho=1000$
the potentials have been multiplied by the factor 100 and the
currents by the factor 1000.} \label{fig-potmuIV}
\end{figure}
%%%%%%%%%%%%%%%%%%%%%%%%%

Under the conditions formulated in Sect.~II when the inelastic
relaxation can be neglected, $\eta \gg \tau_{\rm e-e}^{-1}$, the
kinetic equations (\ref{kineq-1}) and (\ref{kineq-2}) were solved
numerically for nonequilibrium distributions in both normal and
superconducting islands. Throughout numerical computations we use
the value of the pair-breaking parameter $\Delta /\gamma = 1000$.
For nonequilibrium states, the bath temperature was fixed at a
value much lower than the gap, $T_{\rm bath}=(0.1/ 1.764)\,\Delta
$. This corresponds to $T=0.1\, T_c$ if the gap coincides with the
zero-temperature BCS value $\Delta _0 =1.764\,T_c$.
%%%%%%%%%%%%%%%%%%%%%%%%
\begin{figure}[t]
\centerline{\includegraphics[width=1.00\linewidth]{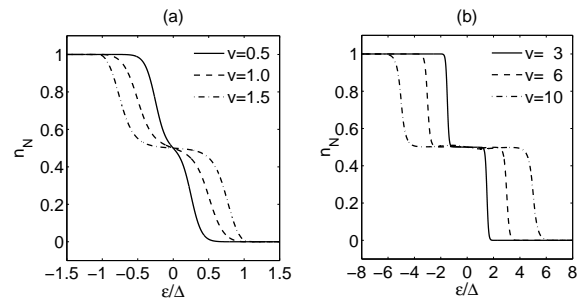}}
\caption{Nonequilibrium normal-island distributions for
$\rho=1000$. We put $v\equiv eV/\Delta$ here and in
Figs.~\protect\ref{fig-fNRmid}--\protect\ref{fig-fSReven} below.}
\label{fig-fNRbig}
\end{figure}
%%%%%%%%%%%%%%%%%%%%%%%%%%

The distribution functions in both N and S islands are in turn
determined by the chemical potential of superconducting islands
$\mu_S$ which was calculated self-consistently using the current
conservation as discussed in the previous section. Chemical
potential and the current--voltage curves are shown in
Fig.~\ref{fig-potmuIV}.

We find three qualitatively distinct cases characterized by the
ratio $\rho\equiv R_{\rm out}/R_{\rm in}$. For the ratio as large
as $\rho=1000$, the distribution in the normal island, $n_N$, is
shown in Fig.~\ref{fig-fNRbig} for several values of $V$. Under
biasing $V$, the distribution is driven into nonequilibrium. At
first, this is seen only as a slight deviation from the Fermi
function $n_F$ but for voltages $V/\Delta \gtrsim 1$, the
distribution $n_N$ assumes the characteristic step-like profile.
This behavior of $n_N$ can be explained as follows. For very high
ratios $\rho$, the superconducting chemical potential is small:
$\mu _S/eV \ll 1$ [compare with Fig.~\ref{fig-potmuIV} (a)]. The
kinetic equations (\ref{kineq-1}), (\ref{kineq-2}) yield then
$f_1(N)\approx f_1(S)\approx f_1(N_1) $. Since $f_2(N)=0$, one has
$2n_N=1-f_1(N)=1-f_1(N_1)$. The external leads $N_1$ and $N_2$ are
in equilibrium such that $f_1(N_1)=-n_F(\epsilon
-eV)+n_F(-\epsilon -eV)$. Thus, $n_N=[n_F(\epsilon -eV)
+n_F(\epsilon +eV)]/2 $. Since $f_2(S)$ is small due to the high
resistivity ratio, one has also $n_S\approx n_N$ for the
superconducting island.

\begin{figure}[t]
\centerline{\includegraphics[width=1.00\linewidth]{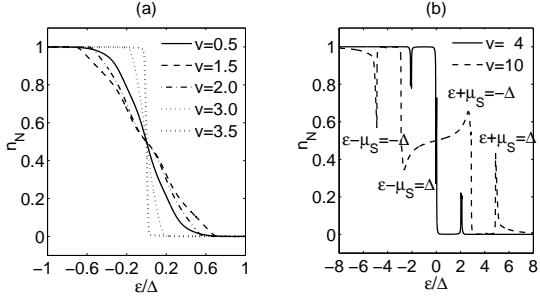}}
\caption{Nonequilibrium normal metal distributions for $\rho=1$.}
\label{fig-fNRmid}
\end{figure}

\begin{figure}[t]
\centerline{\includegraphics[width=1.00\linewidth]{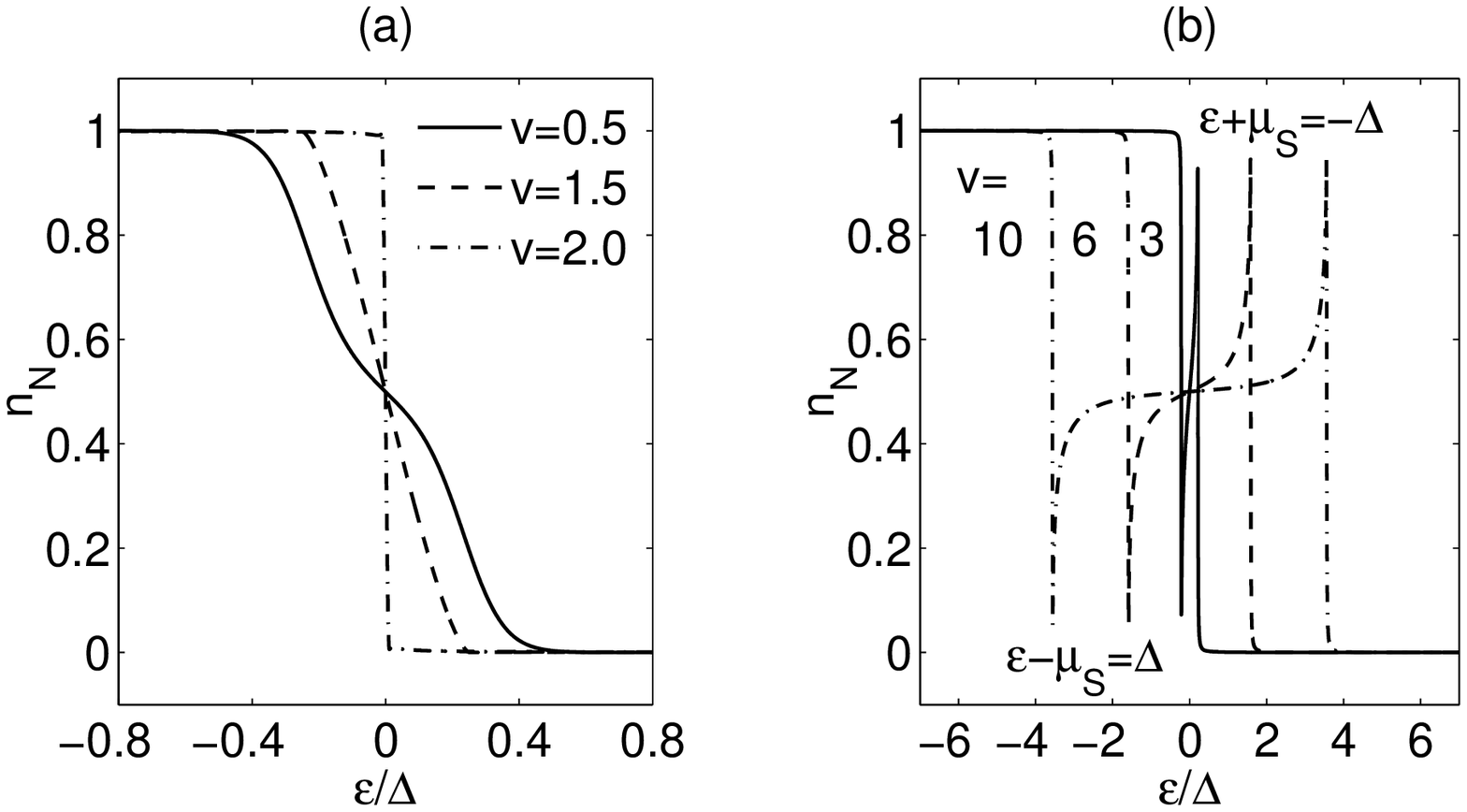}}
\caption{Nonequilibrium normal metal distributions for
$\rho=1/1000$.} \label{fig-fNRsmall}
\end{figure}

\begin{figure}[t]
\centerline{\includegraphics[width=0.5\linewidth]{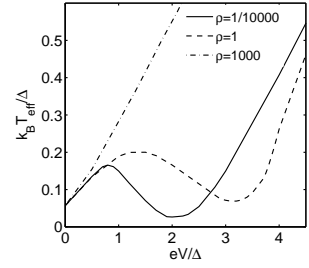}}
\caption{Effective temperature $T_{\rm eff}$ of
Eq.~\eqref{eq-Teff}.} \label{fig-Teff}
\end{figure}

The distributions for the ratios $\rho=1$ and
$\rho=1/1000$ are shown in
Figs.~\ref{fig-fNRmid} and \ref{fig-fNRsmall}. They also change
their shape strongly above a certain voltage. The changes
correspond to the sharp rise in the electric current through the
junction, Fig.~\ref{fig-potmuIV}(b). The origin of this behavior
is discussed in connection with the charge imbalance, see next
subsection. The distribution functions of the N island show a
cooling behavior: The distribution becomes very steep, thus
corresponding to a low effective electron temperature at such bias
voltages when the chemical potential difference across the
superconductor/normal-island junction is $|\mu_S|\approx \Delta$.
One can define an effective temperature through \cite{Pekola2}
\begin{equation}
k_BT_{\rm eff}=\frac{\sqrt{6}}{\pi}
\sqrt{\int_0^\infty\left[1-f_{1,\epsilon}(N)\right]\epsilon\,
d\epsilon } \quad . \label{eq-Teff}
\end{equation}
keeping in mind that $f_2(N)=0$ for the center N island. This
definition gives the actual electron temperature in
(quasi)equilibrium. The effective temperatures have minima for
$\rho\lesssim 1$ as seen in Fig.~\ref{fig-Teff}.

%%%%%%%%%%%%%%%%%%%%%%%
%%%%%%%%%%%%%%%%%%%
%%%%%%%%%%%%%%%%%%%%%
%%%%%%%%%%%%%%%%%%%%%%%
%%%%%%%%%%%%%%%%%%%%%%%%%
%%%%%%%%%%%%%%%%%%%%%%%%%%%%

For a good contact between the outer normal electrodes and the
superconductors S, i.e. when $\rho=1/1000$, there is no deviation
between the superconductor distribution function $n_S$ and that in
the normal reservoir, a Fermi function with $\varphi_{N_1}=V/2$.
This suggests that, for small $\rho$, the considered structure is
similar to a SINIS system with superconducting reservoirs. The
distribution in the central N island for low $\rho$ is shown in
Fig.~\ref{fig-fNRsmall}. It resembles the distribution found for a
SINIS structure \cite{Giazotto}.

%%%%%%%%%%%%%%%%%%%%%%%%%%
\begin{figure}[t]
\centerline{\includegraphics[width=1.00\linewidth]{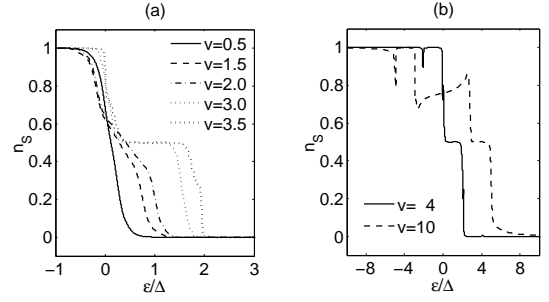}}
\caption{Nonequilibrium superconductor distributions for various
bias voltages when  $\rho=1$.}
\label{fig-fSReven}
\end{figure}
%%%%%%%%%%%%%%%%%%%%%%%%%

When the ratio $\rho$ increases, the distribution in the
superconductors, $n_S$, deviates from the Fermi function as shown
in Fig.~\ref{fig-fSReven}. Nonequilibrium distribution in the
superconducting regions on both sides of the central normal island
drives the state in the N island yet further from equilibrium, see
Fig.~\ref{fig-fNRmid}. For larger $\rho$ the cooling behavior
becomes less pronounced and finally disappears, see
Fig.~\ref{fig-Teff}.

For low and intermediate values of $\rho$, we
observe novel features of highly nonequilibrium distributions both
in the central normal and in the side superconducting islands. For
small $\rho=1/1000$, peaks in the energy
distribution appear at energies $\pm \epsilon = \Delta + \mu_S$
(see Fig.~\ref{fig-fNRsmall}). In addition to these, new peaks
appear in Fig.~\ref{fig-fNRmid} for larger $\rho= 1$ at energies
$\pm \epsilon = -\Delta + \mu_S$, for voltages
considerably exceeding $\Delta /e$. Both sets of peaks come as a
result of recursion from singularities in $g_\epsilon$ and
$F_\epsilon$ at $\epsilon =\pm \Delta$ in the kinetic equations.
The new peaks at $\pm \epsilon = -\Delta + \mu_S$ are present as
long as the distribution functions of the two reservoirs differ at
the corresponding energies and appear as a result of suppression
of the distribution function due to a large factor $(\Delta/\gamma
) F_{\epsilon \pm \mu_S}$ in the sub-gap region in
Eqs.~(\ref{M-trace1}) and (\ref{M-trace2}). Thus, the requirement
of the new peaks is roughly $\Delta -\mu_S<eV/2$ which can be
fulfilled if $\mu_S\ne -eV/2$ as is the case for intermediate
values of $\rho$.

\subsubsection{Charge imbalance}

As mentioned above, the distribution function of the center N
island suffers a drastic change above a certain voltage. This
change coincides with the upturn of the current as a function of
the applied voltage in Fig.~\ref{fig-potmuIV}(b) and is
accompanied by a deviation of the chemical potential $\mu _S$ from
the electric potential $-e\varphi _S$ in the adjacent
superconductor as determined by Eq.~(\ref{charge1}). In
equilibrium, their difference $\Phi =0$. In nonequilibrium, a
difference between $\mu _S$ and $-e\varphi _S$ appears according
to Fig.~\ref{fig-potmuIV}(a). The singularities appear when the
chemical potential difference between the superconductor and one
of the contacting normal conductors approaches $\Delta$. This
corresponds to $eV \sim 2\Delta$ for a large mismatch between
$R_{\rm in}$ and $R_{\rm out}$ or to $eV/2 \sim 2\Delta$ for
$R_{\rm in}=R_{\rm out}$. To measure potentials $\varphi _S$, a
capacitive connection would be required, in addition to the usual
resistive connection only capable of detecting $\mu_S$.

For a good contact between the outer normal electrodes and the
superconductors S, i.e. when $\rho=1/1000$, there is no deviation
between $-e\varphi_S$ and $\mu_S$. This can be understood by
considering the function $n_S$, which coincides essentially with a
Fermi function shifted by $\mu_S\approx -e \varphi_{N_1}=-eV/2$.
For a Fermi function, $n_\epsilon =1-n_{-\epsilon}$, and the term
in the brackets in Eq.~(\ref{charge1}) vanishes.

\subsubsection{Gap instability}

%%%%%%%%%%%%%%%%%%%%%%%%%%
\begin{figure}[t]
\centerline{\includegraphics[width=1.00\linewidth]{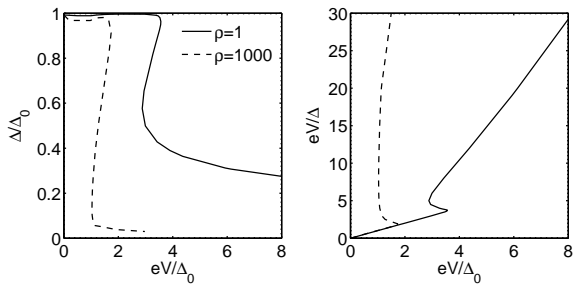}}
\caption{(a) Gap $\Delta$ as a function of bias voltage for two
resistance ratios $\rho=1000$ (dashed line) and
$\rho=1$ (solid line). Both quantities are
normalized to zero-temperature BCS gap $\Delta_0$. (b) Conversion
between $eV/\Delta$ and $eV/\Delta_0$ for $\rho=1000$ (dashed line)
and $\rho=1$ (solid
line).} \label{fig-gap}
\end{figure}
%%%%%%%%%%%%%%%%%%%%%%%

In a nonequilibrium state, the gap function $\Delta$ has to be
calculated self-consistently using Eq.~(\ref{eq-Delta}). Employing
the equation for the critical temperature $T_c$,
\begin{equation}
\frac1\lambda=\int_0^{E_c}\tanh \frac{\epsilon}{2T_c}\,
\frac{d\epsilon}{\epsilon},
\end{equation}
one can exclude the interaction constant in favor of $T_c$. For a
low resistance ratio $\rho=1/1000$, the gap does not change
considerably, $\Delta\approx\Delta_0$ for all $V$, $\Delta _0
=1.764\, T_c$ being the zero-temperature BCS gap. However, for a
higher resistance ratio, the nonequilibrium energy gap is modified
dramatically as shown in Fig.~\ref{fig-gap}(a). One observes a
drastic change in the gap for voltages coinciding with those where
the change in the distribution is seen. The energy gap becomes a
multi-valued function which implies hysteretic behavior
accompanied by jumps of $\Delta$ at the corresponding voltages.
For a very poor contact between the superconducting islands and
the outer normal reservoirs, i.e. for high $\rho$ when deviation
from equilibrium is the largest, the gap function jumps down to
very small values and superconductivity is nearly destroyed. For a
lower tunnel resistance ratio $\rho$, the gap decrease is not so
huge and superconductivity is less suppressed. Note that with
respect to the order parameter magnitude, the bath temperature
chosen for our calculations can be considered as zero. Indeed, the
temperature was set much lower than $\Delta$ while the relevant
energy scales for the distribution function are determined by the
applied voltage and by $\Delta$ itself. Thus the thermal effects
on the gap are negligible.

The predicted suppression of superconductivity in a nonequilibrium
superconductor placed into a tunnel contact with a {\it
nonequilibrium normal-metal electrode} contrasts to the
superconductivity enhancement observed in tunnel SIS$^\prime$IS
structures \cite{SCenhancement1,SCenhancement2,HeslingaKlapwijk}
where the nonequilibrium superconductor is in contact with {\it
equilibrium superconducting} electrodes.

Note that since in our calculations we normalize all the values
with the dimensions of energy (like voltage, excitation energy,
temperature) to the real gap magnitude $\Delta$, we need  to
rescale the real voltage to its relative magnitude. Conversion
between the relative, $V/\Delta$, and the real voltage normalized
to the BCS gap, $V/\Delta_0$, is provided by Fig.~\ref{fig-gap}(b)
where the graphs are given for $\rho=1000$ and $\rho=1$. As seen
from Fig.~\ref{fig-gap}(b), for $\rho=1$ high relative voltages
$eV/\Delta$ can be achieved for comparatively low absolute voltage
values $eV/\Delta _0$.

\subsubsection{Visualization}

The peaks in the distribution of N and S islands can be monitored
by measuring the differential conductance of a probe tunnel SINIS
or SIS$^\prime$IS junction attached to the island in question. Let
us consider a SINIS probe junction attached to the central N
island as in Ref.~\onlinecite{Pekola2}. The distribution function
in the N island is not modified by the measuring current if the
tunnel resistance of the probe junction satisfies $R_{S_PN}\gg
R_{\rm in} +R_{\rm out}$. The current through the probe junction
is
\begin{eqnarray*}
I(N\rightarrow S_P)&=&-\frac{1}{eR_{S_PN}}\int_{-\infty}^{\infty}
g_{\epsilon}(S_P) \\
&& \times \left[n_F(\epsilon)-n_N(\epsilon +eV_P/2) \right]\,
d\epsilon, \label{e-current-P/NIS}
\end{eqnarray*}
where the bulk superconducting probe electrode S$_P$ is assumed to
be in equilibrium with a potential $\varphi_P$ so that $\mu
_{S_P}=-e\varphi _P=-eV_P/2$ where $V_P$ is the voltage between
the two probe electrodes. The energy gap $\Delta_P$ in the probe
electrodes is assumed to have the magnitude corresponding to the
BCS value for $T=T_{\rm bath}$. As $T_{\rm bath}\approx 0.1\,
T_c$, the gap $\Delta _P$ is very close to $\Delta_0$. In
addition, we set $\gamma_P=\gamma$ for the depairing rate in the
probe electrodes. The differential conductance becomes
\begin{equation}
\frac{dI}{dV_P}=\frac{1}{2R_{S_PN}}\int_{-\infty}^{\infty}
 g_{\epsilon}(S_P)\frac{dn_N(\epsilon +eV_P/2)}{d\epsilon}\,
 d\epsilon \ .
\label{dIdVP}
\end{equation}
Due to the peaks in $g_\epsilon (S_P)$ at $\epsilon =\pm \Delta
_P$, the differential conductance should reproduce the peaks in
the distribution function $n_N$ at the probe voltages satisfying
$\epsilon \pm \Delta _P=V_P/2$.

%%%%%%%%%%%%%%%%%%%%%%%%%%%%%%
\begin{figure}[t]
\centerline{\includegraphics[width=1.00\linewidth]{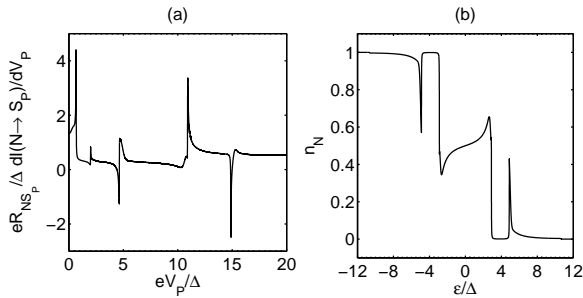}}
\caption{Differential conductance for the probe junction (a) and
the corresponding quasiparticle energy distribution in the
normal-metal island (b) when $\rho=1,
eV/\Delta=10$.} \label{fig-diffcond10}
\end{figure}
%%%%%%%%%%%%%%%%%%%%%%%%%%%
%%%%%%%%%%%%%%%%%%%%%%%%%%%%%
%\begin{figure}[htbp]
%\centerline{\includegraphics[width=0.95\linewidth]{diffcondR1V20-1.eps}}
%\caption{Differential conductance for the probe junction when
%$\rho=1, eV/\Delta=20$ (a) and the corresponding
%quasiparticle energy distribution in the normal metal island (b).}
%\label{fig-diffcond20}
%\end{figure}
%%%%%%%%%%%%%%%%%%%%%%%%%%%%

The differential conductance $dI(N\rightarrow S_P)/dV_P$ together
with the corresponding distributions in the central normal island
for $eV/\Delta=10$ are shown in Fig.~\ref{fig-diffcond10} for the
ratio $\rho=1$. The peaks in the distribution in
Fig.~\ref{fig-diffcond10}(b) are located at
$\epsilon=\pm\Delta-\mu_S$, i.e., $\epsilon=2.89\,\Delta$ and
$\epsilon=4.89\,\Delta$. Two more are located at
$\epsilon=\pm\Delta-3\mu_S$ or $\epsilon=10.7\,\Delta$ and
$\epsilon=12.7\,\Delta$ (out of scale in
Fig~\ref{fig-diffcond10}(b)). Comparing these to the locations of
the larger peaks in Fig.~\ref{fig-diffcond10}(a) and using the
$\Delta$-vs-$\Delta _0$ conversion curve of Fig.~\ref{fig-gap}(a)
we see that the peaks in the distribution are indeed reproduced at
the probe voltage $eV_P/\Delta =2\epsilon /\Delta \pm 2\Delta
_P/\Delta $. The smaller peaks in Fig.~\ref{fig-diffcond10}(a)
refer to the less pronounced structure in the distribution not
well-resolved in Fig.~\ref{fig-diffcond10}(b).

%The corresponding peak in Fig.~\ref{fig-diffcond20}(a) is located
%at $V_P/2=1.05\Delta$. The latter also exhibits fine structure not
%visible in Fig.~\ref{fig-diffcond10}(a), but otherwise the pattern
%repeats itself in the two figures.

\subsection{Quasi-equilibrium}

As discussed in Section II, the condition for full nonequilibrium
is $\eta\gg \tau_{\rm inel}^{-1}$, where $\eta$ is defined
according to Eq.~\eqref{eta}. The inelastic relaxation may be
further separated to relaxation caused by electron--electron and
electron--phonon interactions with collision rates $\tau_{\rm
e-e}^{-1}$ and $\tau_{\rm e-ph}^{-1}$, respectively. The
experimental situation in nanoscale heterostructures\cite{Pekola2}
corresponds often to the case $\tau_{\rm e-e}^{-1}> \tau_{\rm
e-ph}^{-1}$. In particular, the limit of low coupling to the heat
bath in S and/or N islands is frequently realized when the tunnel
injection rate is intermediate between the electron--phonon and
electron--electron relaxation rates, $\tau_{\rm e-ph}^{-1}\ll \eta
\ll \tau_{\rm e-e}^{-1}$. We refer to this case as
quasi-equilibrium. While the near absence of electron--phonon
interactions prohibits the quasiparticles from coupling to the
lattice, the rate of electron--electron scattering is high enough
for the quasiparticles to assume a Fermi distribution with certain
electron temperature. We have studied the cooling performance of
our NISINISIN heterostructure in the quasi-equilibrium limit
looking at the electron temperature of the central N island.

We note that the simple expressions for the regular Green
functions of the form of Eqs.~(\ref{gRA}) and (\ref{fRA}) are not
applicable in the strict sense when the inelastic relaxation
dominates. To find the exact expressions for the regular Green
functions one has to solve the Eilenberger equations
(\ref{Eilen-eq1}), (\ref{Eilen-eq2}) with the proper inelastic
collision integrals. However, to simplify our problem, we model
the pair breaking effects of inelastic relaxation by an effective
pair-breaking rate $\gamma$ in Eqs.~(\ref{gRA}) and (\ref{fRA}) in
the same way as for the tunnel limit described in the previous
sections. This approximation is frequently used in practical
calculations. Here we put $\Delta /\gamma =1000$ as above.

%%%%%%%%%%%%%%%%%%%%%%%%%%%%
\begin{figure}[t]
\centerline{\includegraphics[width=1.00\linewidth]{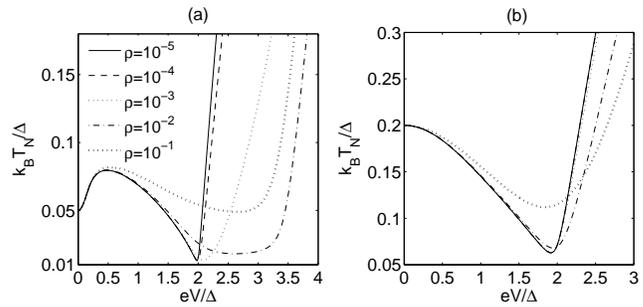}}
\caption{Temperature in the normal-metal island as a function of
bias voltage for bath temperatures 0.05$\,\Delta$ (a) and
0.2$\,\Delta$ (b) when the system is quasiequilibrium.}
\label{fig-qeqTN}
\end{figure}
%%%%%%%%%%%%%%%%%%%%%%%%%%%%

%%%%%%%%%%%%%%%%%%%%%%%%%%%%
\begin{figure}[htbp]
\centerline{\includegraphics[width=1.00\linewidth]{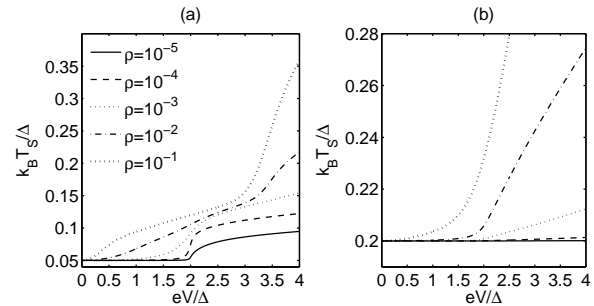}}
\caption{Temperature in the superconductor as a function of bias
voltage for bath temperatures 0.05$\,\Delta$ (a) and 0.2$\,\Delta$
(b) when the system is quasiequilibrium.} \label{fig-qeqTS}
\end{figure}
%%%%%%%%%%%%%%%%%%%%%%%%%%%%

Applying the tunnelling model shows that depending on the
configuration of the quasiparticle traps, i.e. the ratio of outer
and inner junction resistances $\rho$, effective cooling of the
normal-metal island can be achieved as demonstrated also in
Ref.~\onlinecite{Pekola2}. The temperatures of the central N
island and of the contacting S island are shown in
Figs.~\ref{fig-qeqTN} and \ref{fig-qeqTS}, respectively. The
temperature of N island indeed displays a minimum below bath
temperature when the ratio $\rho$ is small. However, for large
$\rho$, the temperature monotonously rises above the bath
temperature with an increasing voltage. As can be deduced from
Figs.~\ref{fig-qeqTN} and \ref{fig-qeqTS}, the cooling effect is
not attributed to the presence of two additional normal metal
reservoirs. On the contrary, smallest ratio $\rho$, which
corresponds to the strongest cooling, is seen to lead to almost
constant $T_S=T_{\rm bath}$ as it would be the case for pure
superconducting reservoirs. Another remark concerns the cooling
efficiency of a NISINISIN configuration as $T_{\rm bath}$ is
lowered. Indeed, for $T_{\rm bath}=0.2\,\Delta$ the temperature
minimum for the depairing parameter $\gamma =\Delta/1000 $ used
for our calculations is roughly $0.063\,\Delta$ while the minimum
is $T_{\rm min}=0.024\,\Delta$ for $T_{\rm bath}=0.15\Delta$, and
it is $T_{\rm min}\approx 0.013\,\Delta$ for $T_{\rm bath}\leq
0.1\, \Delta$. From the numerical results in the case $1 \gg \rho
\gtrsim \gamma/\Delta$, we find that the minimum achieved
temperature follows $T_{\rm min}/T_c \approx 0.24 \rho ^\zeta$
where $\zeta \approx 0.5$. For smaller $R_{\rm out}$, the minimum
temperature $T_{\rm min}$ is determined by the inverse proximity
effect described by the depairing parameter $\gamma$ (see
Ref.~\onlinecite{Pekola2}). Combining both the superconductor
heating due to a finite $R_{\rm out}$ and the effects of
depairing, we can write an approximate formula for the minimum
temperature for relatively low bath temperatures,
\begin{equation}\label{eq:tmin}
T_{\rm min}/T_c =2.5 \left(\gamma /\Delta \right)^{2/3}+0.24 \rho
^{1/2} \ .
\end{equation}
For $\rho  \ll 1$, the depairing rate $\gamma$ is limited by
$R_{\rm out}$, i.e., $\gamma = 1/(4 \nu_S e^2 \Omega_S R_{\rm
out}) = (r_S/2 R_{\rm out})E_T$, where $E_T=D/L^2$ is the Thouless
energy in the superconducting island with length $L$ and $r_S$ is
its resistance. Substituting this into Eq.~\eqref{eq:tmin}, we
find that the minimum temperature is optimized with
\begin{equation}\label{eq:ropt}
\rho \approx 6.4 \left[(E_T/\Delta)( r_S/R_{\rm in}) \right]^{4/7}
\ .
\end{equation}
Note, however, that Eq.~\eqref{eq:ropt} is valid provided
$r_S/R_{\rm in} \ll \rho \ll 1$.

%%%%%%%%%%%%%%%%%%%%%%%%%%%
\begin{figure}[t]
\centerline{\includegraphics[width=1.00\linewidth]{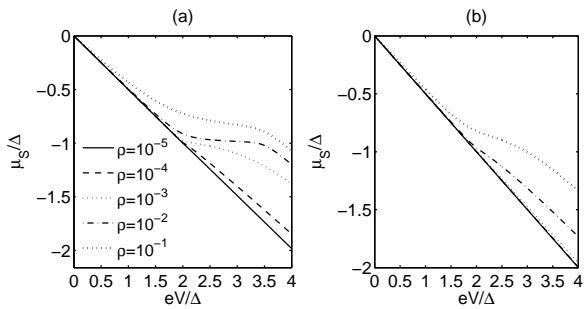}}
\caption{Chemical potential of the superconductor as a function of
bias voltage for bath temperatures 0.05$\,\Delta$ (a) and
0.2$\,\Delta$ (b) when the system is in quasiequilibrium.}
\label{fig-qeqmu}
\end{figure}
%%%%%%%%%%%%%%%%%%%%%%%%%%%%

%%%%%%%%%%%%%%%%%%%%%%%%%%%%
\begin{figure}[t]
\centerline{\includegraphics[width=1.00\linewidth]{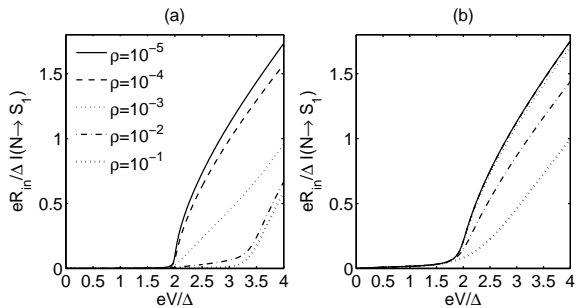}}
\caption{Electric current between S and N islands as a function of
bias voltage for bath temperatures 0.05$\,\Delta$ (a) and
0.2$\,\Delta$ (b) when the system is in quasiequilibrium.}
\label{fig-qeqI}
\end{figure}
%%%%%%%%%%%%%%%%%%%%%%%%%%%%

%%%%%%%%%%%%%%%%%%%%%%%%%%%%
\begin{figure}[t]
\centerline{\includegraphics[width=1.00\linewidth]{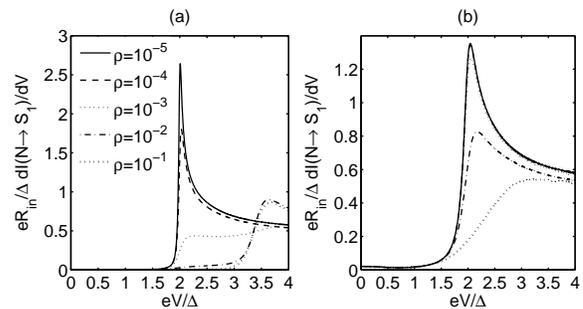}}
\caption{Differential conductance as a function of bias voltage
for bath temperatures 0.05$\,\Delta$ (a) and 0.2$\,\Delta$ (b)
when the system is in quasiequilibrium.} \label{fig-qeqG}
\end{figure}
%%%%%%%%%%%%%%%%%%%%%%%%%%%%

The sharp rise in $T_N$ and $T_S$ occurs generally around
$V=2\,\Delta$ but for larger values of $\rho$,
when also $T_{\rm bath}\rightarrow T_{\rm min}$, the upturn shifts
towards higher voltages $V$. This is because the upturn is
determined by the condition $-\mu_S\approx \Delta$ rather than
$V/2=\Delta$ as can be seen by comparing Figs.~\ref{fig-qeqTN} and
\ref{fig-qeqTS} to Fig.~\ref{fig-qeqmu}. For increasing bath
temperature, though, this trend is smeared and disappears. The
voltages corresponding to the temperature rise are also seen in
the IV-curves in Fig.~\ref{fig-qeqI} and in the differential
conductance, Fig.~\ref{fig-qeqG}.

\section{Discussion}
\subsection{Electron cooling}

According to our results, the electron cooling is the most
effective when the outer resistance is low $\rho\ll 1$. In fact,
both the effective temperature and the
distribution function in the superconducting region almost
coincide with those in the bath for such voltages that yield
$|\mu_s| <\Delta$ especially for very low ratios of $\rho$. When the
ratio $\rho$ is low, a good
contact between the inner superconducting region and the outer
electrode makes the distributions in these two regions not so much
different from each other, thus decreasing the role of the extra
junction. This conclusion is valid only within the tunneling
approximation. When the contact between the outer electrodes and
superconducting islands are more transparent, cooling properties
of the device are affected by the inverse proximity effect from
the external normal leads.

For larger ratios $\rho$, the extra junction
prevents the state of the superconducting region from reaching
equilibrium, thus reducing the cooling power of the entire
structure. Moreover, this limit has another disadvantage as far as
the cooling performance is concerned: For larger voltages when
$eV$ approaches $2\Delta$, one expects a suppression of
superconductivity in the $S$ regions down to lower values of
$\Delta$ and thus the cooler would become even less effective.
This suggests that the cooling performance of an NISINISIN
structure cannot be improved essentially by an extra tunnel
junction as compared to that of a simple SINIS structure. However,
the presence of the quasiparticle traps helps to practically
realize the superconducting reservoirs by thermalizing them
quickly to an object with a high thermal conductance.

\subsection{Nonequilibrium distribution}

Nonequilibrium distribution formed in the superconducting islands
for high values of $\rho$ results in yet
stronger deviation from equilibrium in the central normal island.
As seen from Figs.~\ref{fig-fNRmid} and \ref{fig-fNRsmall}, the
distribution function in the N island is characterized by peaks at
energies $\pm \epsilon = \Delta \pm \mu_s$, and also $\pm \epsilon
= \Delta \pm 3\mu_s$, etc., for voltages considerably exceeding
$\Delta /e$. These peaks are clearly visible in the differential
conductance of a probe tunnel SINIS junction made at the central
normal island, Fig.~\ref{fig-diffcond10}. Simultaneously,
nonequilibrium states in the superconducting region follow the gap
which is strongly reduced as compared to its equilibrium BCS value
$\Delta _0$. The transition into a nonequilibrium state is
accompanied by a jump in the gap magnitude which leads to the jump
in the relative voltage: high values of $eV/\Delta$ can be reached
already for not very large absolute values of voltage
$eV/\Delta_0$. This makes observation of the nonequilibrium states
in N and S regions more easily accessible in experiments.

\acknowledgments

We are thankful to J.P. Pekola for stimulating discussions. TTH
acknowledges funding by the Academy of Finland and the NCCR
Nanoscience.

\end{document}